\documentstyle[prl,aps,multicol,epsf,epsfig]{revtex}
\newcommand{\be}{\begin{eqnarray}}
\newcommand{\ee}{\end{eqnarray}}
\newcommand{\la}{\langle}
\newcommand{\ra}{\rangle}
\newcommand{\no}{\nonumber}
\newcommand{\bmath}{\begin{mathletters}} 
\newcommand{\emath}{\end{mathletters}}
\newcommand{\cl}{{\cal L}}
\newcommand{\alp}{\alpha}
\newcommand{\h}{\hat}

\begin{document}

\title
{Dispersive Lineshape Theory}

\author{YounJoon Jung, Eli Barkai, and Robert J. Silbey}

\address
{Department of Chemistry, Massachusetts Institute of Technology \\
Cambridge, MA 01239}

\date{\today}

\maketitle

\begin{abstract}
Motivated by recent experiments 
we consider a stochastic lineshape theory for the case when the 
underlying process obeys power-law statistics, based on a generalized Anderson-Kubo 
oscillator model. We derive an analytical expression for the lineshape 
and find rich type of behaviors when compared 
with the standard theory, for example,  
new type of resonances and narrowing phenomena.
We show that the lineshape is extremely 
sensitive to the way the system is prepared at time $t=0$ and discuss the problem of stationarity. 

\vspace{0.1in}
\end{abstract}


\begin{multicols}{2}
\narrowtext 
Since its first introduction by Anderson and Kubo(AK)
\cite{anderson-jpsj-54,kubo-jpsj-54}
lineshape theory based on a stochastic approach has 
had wide applications 
in condensed phase spectroscopy ranging 
from magnetic resonance spectroscopy\cite{anderson-jpsj-54,kubo-jpsj-54}
to the recently developed single molecule spectroscopy
\cite{reilly-jcp-94,geva-jpcb-97,barkai-prl-00}. 
The AK approach 
as well as other standard approaches
(e.~g.~Bloch equation) is based on the Markovian assumption. 
These approaches have been applied with great success mainly to ensemble 
averaged measurements. 
However, single molecule spectroscopy\cite{moerner-prl-89,orrit-prl-90} 
has revealed that in some cases the underlying dynamics in the condensed phase is 
highly non-Markovian
\cite{zumofen-cpl-94,wang-prl-95,lu-science-98,berezhkovskii-jcp-99,chernyak-jcp-99,gangopadhyay-cpl-98,kuno-jcp-00,shimizu-00}.

In recent fluorescence intermittency studies on single quantum dot systems  
\cite{kuno-jcp-00,shimizu-00}
it was found that single quantum dots undergo transitions between bright and dark states during the measurement, 
and the sojourn times of both bright and dark states are distributed according to power-law, $t^{-3/2}$, 
in contrast to an exponential as in the Poisson process. 
Here we are interested in the stochastic lineshape theory for 
such non-Markovian processes with emphasis on 
power-law behavior of the underlying dynamics. 
The introduction of a power-law process 
in the lineshape problem appears quite naturally. 
Consider the following model as an example. 
The static chromophore $C$ at the origin interacts with 
the single perturber $P$ via a short range 
interaction depending on $r$, the distance between $C$ and $P$. 
$C$ absorbs light with a frequency $\omega_0$ when $r<r_c$ and 
with a frequency $\omega_0'\neq\omega_0$ when $r>r_c$. 
When $P$ moves away from $C$ it performs a random walk 
(assumed one dimensional for simplicity). 
Since the probability density of the first return time in the one dimensional random walk 
follows  $t^{-3/2}$\cite{weiss-random-94,bouchaud-physrep-90}, 
the chromophore absorbs light at frequency $\omega_0'$ 
with times distributed according to a power-law distribution. 

An important issue in a power-law stochastic process 
is {\it stationarity}, which is of concern 
due to a very broad temporal distribution for underlying processes. 
The significant difference between stationary and nonstationary cases 
has manifested itself in many physical problems, e.~g.~transport properties in disordered materials\cite{lax-prl-77} 
and power-spectra in chaotic systems\cite{zumofen-physd-93}.
We take into account the stationarity issue fully  
and show that the lineshape is a very sensitive measure of 
stationarity when the underlying dynamics obeys power-law statistics. 
  
The stochastic lineshape theory is based on the equation of motion for 
the transition dipole,
$\dot{\mu}(t)=i\omega(t)\mu(t),$
where $\omega(t)$ is the stochastic frequency of the 
oscillator.
The dynamical quantity which determines the lineshape 
is the relaxation function 
$\Phi(t,t_0)=\la {\mu(t)\mu(t_0)^{*}}\ra,$
where the average is taken over 
all the possible realizations of the underlying stochastic process
and we have set $\la {|\mu(t_0)|^2}\ra=1$.
From the equation of motion we can calculate the relaxation function as 
$\Phi(t,t_0)=\left \la {\exp\left(i\int_{t_0}^{t}d\tau\omega(\tau)\right)}\right\ra$.
When the process is assumed to be stationary, 
$\Phi(t,t_0)=$$\Phi(t-t_0)$, then the lineshape  $I(\omega)$ can be calculated 
as the Fourier transform of the relaxation function by making use 
of the Wiener-Khintchine(WK) theorem\cite{kubo-statphys2-91},
\be
I(\omega)={1\over 2\pi} \int_{-\infty}^{\infty}dt e^{-i\omega t} 
          \Phi(t) 
         ={1\over \pi} {\rm Re} \h{\Phi}(i\omega+\epsilon), \label{line}
\ee 
where the symmetry of $\Phi(t)$, $\Phi(-t)=\Phi^{*}(t)$, has been used 
and the Laplace transform of $z(t)$ denoted by 
$\hat{z}(s)=\cl\{z(t)\}$ and $\epsilon\rightarrow 0^+$.

We assume that the underlying process is a renewal process as in the AK approach. 
To make the model as simple as possible, we consider a two state model. 
The transition frequency $\omega(t)$ of the chromophore  
can take the value of either $-\omega_0$ or $+\omega_0$ 
depending on the perturber state, $|+\ra$ or $|-\ra$, respectively.   
Each alternating path between the states $|+\ra$ and $|-\ra$ of the perturber 
leads to a stochastic realization of chromophore frequency modulation, and 
it is characterized by a sequence of sojourn times in the states $|+\ra$ and $|-\ra$.
The sojourn times in the states $|\pm\ra$, $t_{\pm}$, 
are assumed as mutually independent, identically distributed random variables 
described by the probability density functions(PDFs), $h_\pm(t_\pm)$.
The original AK process amounts to the exponential sojourn time PDF.
We do not assume any specific functional forms for 
the sojourn time probability densities 
from the beginning, but are mainly interested in the process where 
the sojourn times are distributed with long time power-law tails, $t^{-(1+\alpha)}$
($\alpha>0$). 

Assuming stationarity, which is justified if the 
process has been going on for long times before the beginning of observation, 
the sojourn time PDFs for the {\it first} transition event is given by
$f_{\pm}(t_\pm)$ [different than $h_{\pm}(t_\pm)$] 
by a standard argument\cite{feller-prob-57},
\be
f_{\pm}(t_\pm)=\tau_{\pm}^{-1}
\int_{t_\pm}^{\infty}d\tau h_{\pm}(\tau), \label{ht}
\ee
where the mean sojourn time $\tau_{\pm}$$=\int_0^{\infty}dt t h_{\pm}(t)$
is assumed to be finite. 
When $\tau_\pm\rightarrow\infty$ the concept of stationarity breaks down. 
For the Poissonian case $h_{\pm}(t_\pm)=\tau_{\pm}^{-1}\exp(-t_\pm/\tau_{\pm})$, 
so we have $f_{\pm}=h_{\pm}$. 
Therefore, stationarity is naturally satisfied in the Poissonian 
process. 
However, the non-Poissonian process in which we are interested 
will not be stationary if we simply 
set  $f_{\pm}=h_{\pm}$, and the WK theorem therefore does not hold. 

The conditional relaxation functions $\Phi_{ij}$$(i,j=+,-)$ are defined 
over the stochastic paths that start from the state $|i\ra$ at time 0 
and end with the state $|j\ra$ at time $t$.
Here we give a sketch of the derivation for $\Phi_{++}$ by 
summing all the possible stochastic paths that start from and end at the state 
$|+\ra$. 
Along a particular path if no transition is ever made until $t$, 
then the contribution of this path to $\Phi_{++}$ 
is given by $F_{+}(t_+)e^{-i\omega_0 t_+}$ in the time domain, where 
$F_{\pm}(t_\pm)=\int_{t_\pm}^{\infty} d\tau f_{\pm}(\tau)$ 
is the survival probability of the states $|\pm\ra$ 
for the first event.
This contribution will amount to $\h{F}_{+}(s+i\omega_0)$ in the Laplace domain. 
The next possible paths are those which make the first transition to the state $|-\ra$ at $t_{1+}$, 
jump back to the state $|+\ra$ after remaining at the state $|-\ra$ 
for time $t_{2-}$, and stay at the state $|+\ra$ until time $t$. 
The contribution of these to $\Phi_{++}(t)$ is given by
$\int_{0}^{\infty}dt_{1+}$$\int_{0}^{\infty}dt_{2-}$$\int_{0}^{\infty}dt_{3+}$
$\times$$f_+(t_{1+})e^{-i\omega_0 t_{1+}}h_-(t_{2-})e^{i\omega_0 t_{2-}}
H_{+}(t_{3+})e^{-i\omega_0 t_{3+}}$ with the constraint $t_{1+}+t_{2-}+t_{3+}=t$, 
and $H_{\pm}(t_\pm)=\int_{t_\pm}^{\infty}d\tau h_{\pm}(\tau)=$$\tau_{\pm}f_\pm(t_\pm)$ 
is the survival probability corresponding to $h_{\pm}(t_\pm)$.
In the Laplace domain this will read as 
$\h{f}_{+}(s+i\omega_0)\h{h}_{-}(s-i\omega_0)\h{H}_{+}(s+i\omega_0)$ 
by the convolution theorem.
Summing all the possible stochastic paths, we have 
\be 
\h\Phi_{++}(s)&=&\h F_{+}+\h f_{+}\h h_{-}(1+\h h_{+}\h h_{-}+
(\h h_{+}\h h_{-})^2+\cdots)\h H_{+} \no \\
&=&\h F_{+}+{\h f_{+}\h h_{-}\h H_{+}
                     \over{1-\h h_{+}\h h_{-}}},
\ee
where $\h x_{\pm}\equiv\h x_{\pm}(s\pm i\omega_0)$ with $\h{x}$ being 
$\h{f}$,$\h{h}$,$\h{F}$, or $\h{H}$.
In a similar way, we have 
\be
\h\Phi_{+-}(s)&=&{\h f_{+}\h H_{-}\over {1-\h h_{+}\h h_{-}}}, 
\ee
and $\h\Phi_{--}(s)$ and $\h\Phi_{-+}(s)$ are obtained from 
$\h\Phi_{++}(s)$ and $\h\Phi_{+-}(s)$ by exchanging $+$ and $-$.

The total relaxation function can be calculated from 
the conditional relaxation functions, 
\be 
\h\Phi(s)=\sum_{i=\pm}\sum_{j=\pm}p_i\h\Phi_{ij}(s),
\label{2relaxS}
\ee
with the initial distribution of the perturber state 
given by $p_\pm=\tau_\pm/\sum_{i=\pm}\tau_i$. 
Then from Eq.~(\ref{line}) the lineshape is given by 
\be 
I(\omega)
={1\over \pi}{\rm Re}&&\left [\left ({p_+\over z_+}+{p_-\over z_-}\right ) - 
{1\over {\tau_{+}+\tau_{-}}}\left ({1\over z_+}- {1\over z_-}\right)^2\right . \no \\
&&\times\left. {(1-\h h_+)(1-\h h_-) \over 1-\h h_+ \h h_-}\right], \label{sline}
\ee 
where $z_{\pm}=i\omega\pm i\omega_0$ and $\h h_{\pm}=\h h_{\pm}(z_{\pm})$, 
and we have expressed $\h{f}_\pm$,$\h{F}_\pm$, and $\h{H}_\pm$ 
in terms of $\h{h}_\pm$. 
This is the final expression of the lineshape function 
for the stochastic oscillator undergoing the 
two state frequency modulation.

It is our aim here to show that the theory exhibits a very 
{\it strong sensitivity} on the choice of PDF for 
the first event. 
This becomes important for experimental situations when it is not always 
clear if the underlying process is stationary or not. 
For this purpose we define a quasi lineshape $I_{NS}(\omega)$ 
by replacing $\h{f}_{\pm}$ with $\h{h}_{\pm}$ in the derivation 
of Eq.~(\ref{sline}), 
\be
I_{NS}(\omega)={1\over \pi} {\rm Re}\left [{1\over{1-\h{h}_{+}\h{h}_{-}}}  
\left\{p_+ \left({1-\h{h}_{+}\over z_{+}}+ \right.\right.\right. \no \\
\left.\left.\left. {\h{h}_{+}(1-\h{h}_{-})\over z_{-}}\right)
+ p_- \left({1-\h{h}_{-}\over z_{-}} + {\h{h}_{-}(1-\h{h}_{+})\over z_{+}}
\right)\right\}\right]. \label{nsline}
\ee 
Note that for Poissonian case $I_{NS}(\omega)=I(\omega)$ and 
for ordinary processes (i.e. when all moments of $h_{\pm}(t)$ exist) 
$I_{NS}(\omega)\approx I(\omega)$. However, as we show here, 
strong sensitivity on the first event is recovered for the dispersive case. 
We emphasize that Eq.~(\ref{nsline}) is not a lineshape 
because the underlying process is not stationary.
In certain physical situations, however, 
the underlying process is not an ongoing process, but 
has been initiated by a measurement itself, for instance, 
the blinking process in the quantum dot experiment mentioned 
before. 
In this case one can calculate the relaxation function 
$\Phi_{NS}(t)\equiv\Phi(t,0)$ for the nonstationary stochastic process 
and $I_{NS}(\omega)$ is obtained as the complex Laplace 
transform of a well-defined nonstationary relaxation function, $\Phi_{NS}(t)$,
$I_{NS}(\omega)\sim\int_{0}^{\infty}dt e^{-i\omega t}\Phi_{NS}(t)$. 

The original AK model is recovered 
from Eq.~(\ref{sline}) by choosing an exponential sojourn time PDF.
\cite{anderson-jpsj-54,kubo-jpsj-54}.  
We first consider the sojourn time PDFs which have finite first moments, 
$\tau_\pm<\infty$, but divergent second moments. 
As a representative of this class, we use the following form, 
\be
h_{\pm}(t_\pm)&=&\left(\tau_\pm^{3}/2\pi t_\pm^5\right)^{1/2}\exp\left(-{\tau_\pm/2t_\pm}\right). 
\label{f3_2}
\ee
In this case, $h_{\pm}(t)$ 
decays as $t^{-5/2}$ at long times, thus the 
first moment exists, but the second moment diverges. 

\begin{figure} 
\epsfxsize=3 in
\epsffile{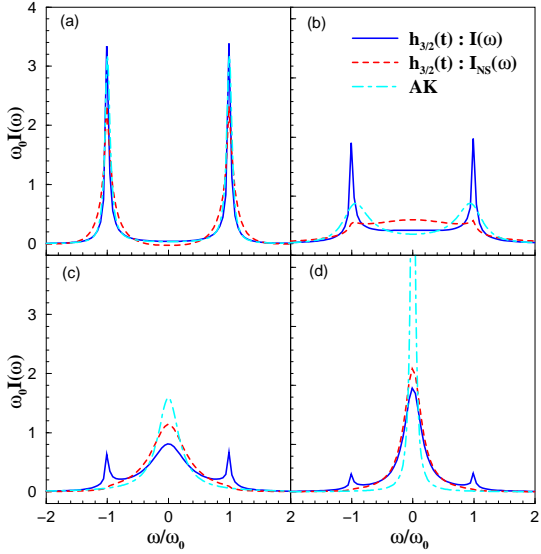}
\caption{
The lineshapes calculated from $h_{3/2}(t)$ 
for the stationary and the nonstationary cases are compared 
with the AK case from slow to fast modulation limits. 
The resonance frequency is chosen as $\omega_0=1$. The 
correlation time is chosen as $\omega_0\tau=20$, $4$, $0.4$, and $0.08$ in (a)-(d),
respectively. 
} \label{fig1}
\end{figure} 

In Fig.~\ref{fig1} we have compared the stationary and 
nonstationary lineshapes for $h_{\pm}(t)$ in Eq.~(\ref{f3_2}) 
and for the AK case. 
We have chosen $h_{3/2}\equiv h_{+}=h_{-}$ by setting $\tau_+=\tau_-=\tau$.
The correlation time of the perturber dynamics  
is varied from slow ($\omega_0\tau\gg 1$) to fast ($\omega_0\tau\ll 1$) modulation 
cases in (a)-(d).
For the AK case, the well known phenomenon of motional narrowing is shown: 
in the slow modulation case we see two peaks at 
$\omega=\pm\omega_0$ while in the fast modulation case 
we observe a single peak at $\omega=0$. 
For the case $h_{3/2}$ 
the stationary and nonstationary 
cases show very different behaviors 
as the correlation time is decreased; thus, 
the first event in the underlying random process 
has a strong effect on the lineshapes. 
In addition, new phenomena are found for the stationary lineshape in 
the fast modulation cases (Fig.~\ref{fig1}(c) and (d)): 
three distinct peaks are observed for the stationary case.

The new peaks we observe in Fig.~\ref{fig1}(c) and (d) at $\omega=\pm\omega_0$ 
for the stationary lineshape result from the 
first event in the stochastic process $\omega(\tau)$. 
The probability for the perturber remaining at the initial state is governed 
by the long time tail in the sojourn time PDF. 
Due to the stationarity condition in Eq.~(\ref{ht}) 
the survival probability for the first event decays 
more slowly for the stationary case($\sim t^{1-\alp}$) than for the 
nonstationary case($\sim t^{-\alp}$), where $\alp=3/2$ in this example.
Therefore, the stationary case effectively 
requires the perturber to remain at the initial state until much longer times 
than the nonstationarity case, resulting in the 
enhanced peaks at $\omega=\pm\omega_0$.
This is why we observe new peaks not present in the standard Poissonian case. 

As the next example we consider the one-sided 
L\'evy density as the sojourn time PDF\cite{bouchaud-physrep-90},
$\h h(s)=\h L_\alp(s/r)=\exp(-(s/r)^\alp)$, 
with $0<\alp<1$, and $r$ being 
a coefficient with an inverse time dimension. 
It is well known that the L\'evy PDF decays 
algebraically at long times $rt\gg 1$,
$L_\alp(t)\sim t^{-(1+\alp)}$, and thus all the moments of $L_\alp(rt)$ 
including the first moment diverge\cite{bouchaud-physrep-90}. 
Therefore, there is no microscopic timescale for this PDF, and 
the form of $f_\pm(t)$ given in Eq.~(\ref{ht}) 
cannot be applied. 
However, in realistic situations, it is unlikely to have power-law behavior 
for an infinitely long time, but rather, it is likely to have a finite 
cut-off time provided by, for example, the lifetime of a molecule. 
Therefore, it is natural to 
introduce a cut-off time $t_c$ such that the algebraic decay is valid during 
time interval $r^{-1}\ll t\ll t_c$. 
We introduce an exponential cut-off function for the convenience of 
an analytical treatment.  
Now the sojourn time PDF is given by 
$
h_{\pm}(t)={\cal{N}}_{\pm} e^{-t/t_c} L_{\alpha}(r_{\pm}t),
$
where ${\cal{N}}_{\pm}$ is the proper normalization constant 
depending on the cut-off time. Then the Laplace domain expression of $h_{\pm}(t)$ 
can be written as 
\be
\h h_{\pm}(s)=\exp\left[(r_\pm t_c)^{-\alp}\left\{1-(1+ s t_c)^\alp\right\}\right]. \label{levyc}
\ee
Then 
$f_{\pm}(t)$ is given from 
Eq.~(\ref{ht}) with the mean being $\tau_\pm=\alpha t_c/(r_\pm t_c)^{\alp}$. 
Note that in the limit  $t_c\rightarrow\infty$ the L\'evy PDF without 
cut-off is recovered as $\h h_\pm (s)=\exp(-(s/r_\pm)^\alp)$ 
and $\tau_{\pm}$ diverges.

\begin{figure} 
\epsfxsize=3 in
\epsffile{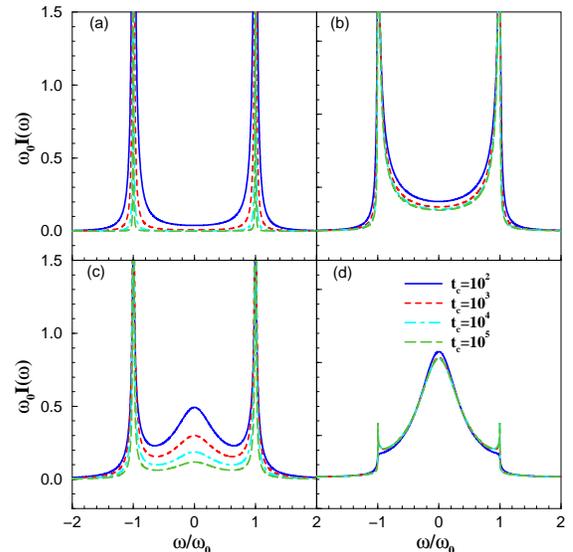} 
\caption{
The lineshapes are calculated with the L\'evy PDF cases with two different 
values of the L\'evy index, $\alp$.  The other parameters are chosen as 
$\omega_0=1$ and $r_{\pm}=1$, except for the cut-off time which is varied.  
(a) $\alp=0.3$, stationary, (b) $\alp=0.3$, nonstationary, 
(c) $\alp=0.8$, stationary, and (d) $\alp=0.8$, nonstationary cases. 
} \label{fig2}
\end{figure} 

In Fig.~\ref{fig2} we have investigated the effect of the cut-off time 
in the L\'evy PDF case both in the stationary and the nonstationary cases. 
When $\alp=0.3$ (Fig.~\ref{fig2} (a) and (b)), 
both the stationary and the nonstationary 
L\'evy lineshapes show distinct peaks at 
$\omega=\pm\omega_0$.
As the cut-off time is increased, there is little change in both lineshapes 
other than becoming narrower. 
When $\alp=0.8$ (Fig.~\ref{fig2} (c) and (d)), 
there appears a new peak near $\omega=0$ 
in addition to the two resonance peaks. 
This is a new type of the narrowing behavior which is absent in the Poissonian 
case in that it is controlled by the power-law index $\alpha$ rather than the 
correlation time $\tau$ (as in the Poissonian case), and is termed 
{\it power-law narrowing}.
Also, as the cut-off time is increased, the central 
peak in the lineshape for the stationary case diminishes while it remains in the 
nonstationary case. This is because in the stationary case, as the cut-off time is 
increased the first event will dominate the probability weight 
in the stochastic paths of the perturber dynamics. The difference between 
the stationary and nonstationary cases is therefore more significant in the L\'evy case 
than in the case $h_{3/2}(t)$ given in Fig.~1. 

To investigate the power-law narrowing 
we consider the limit $t_c\rightarrow\infty$.  
In this limit the stationary lineshape approaches 
two delta functions,
$I(\omega)=p_+\delta(\omega+\omega_0)+p_-\delta(\omega-\omega_0)$,
since the second term in the Eq.~(\ref{line}) vanishes as 
$\tau_\pm\rightarrow\infty$.
The nonstationary case, however, yields in the limit $r_{\pm}\gg|\omega|$,
\be
\lim_{t_c\rightarrow\infty}I_{NS}(\omega)=
{\sin(\pi\alp) \over 2\pi\omega_0}
{2+x+x^{-1}\over \eta x^\alp+(\eta x^{\alp})^{-1}+2\cos(\pi\alp)}, 
\label{nsline1}
\ee 
for $|\omega|<\omega_0$ and vanishes when $|\omega|>\omega_0$. 
This expression has been obtained from Eq.~(\ref{nsline}) 
by taking the small frequency limit of the sojourn time PDF, 
$\h h_{\pm}(s)=1-(s/r_{\pm})^\alp+\cdots$. Note that Eq.~(\ref{nsline1}) 
is not limited to L\'evy PDF, but valid for any PDF with $t^{-(1+\alp)}$ tail ($0<\alp<1$).
Here, $x=(\omega_0+\omega)/(\omega_0-\omega)$ is a dimensionless frequency,  
and $\eta=\lim_{t_c\rightarrow\infty}(p_+/p_-)=(r_-/r_+)^\alp$ 
is the asymmetry parameter. This function shows a very asymmetric 
power law singularities at 
$\omega=\pm\omega_0$, $(\omega_0\pm\omega)^{1-\alpha}$
depending on $\alpha$. 
It is worthwhile to mention that such a strong asymmetric lineshape 
has been encountered in the problem of the X-ray edge absorption of metals
\cite{mahan-manyptl-81}.
In the symmetric case ($\eta=1$) Eq.~(\ref{nsline1}) reduces to a simpler expression, and 
in this case there exists a 
critical value of $\alpha$ below which the lineshape is concave and above which convex 
at $\omega=0$, which is given by $\alpha_c=\cos(\pi\alpha_c/2)=0.5946\cdots$.

\begin{figure} 
\epsfxsize=3 in
\epsfysize=2 in
\epsffile{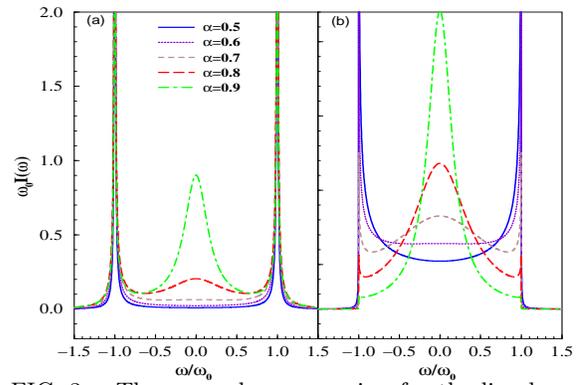} 
\caption{
The power-law narrowing for the lineshape with L\'evy PDF is 
shown as the L\'evy index $\alpha$ is changed. The other parameters are 
chosen as $r_{\pm}=100$ and $t_c=10^4$. (a) stationary and (b) nonstationary cases.
} \label{fig3}
\end{figure}

In Fig.~\ref{fig3}
we have confirmed this finding by plotting the nonstationary lineshapes for the 
L\'evy PDF case with $\eta=1$ and finite, large $r_\pm$ and $t_c$. 
The nonstationary case in Fig.~\ref{fig3} (b) shows the concave-to-convex transition at 
the critical value of $\alpha$ as predicted. 
For the stationary lineshape, similar kind of behavior 
can be observed in Fig.~\ref{fig3} (a), however, $\alpha_c$ now depends on $t_c$. 

In summary, we have generalized the stochastic lineshape 
theory to arbitrary renewal processes. 
Compared with the standard theory, we have found a variety of new 
phenomena in the lineshapes, such as new peaks and narrowing behaviors. 
The issue of the stationarity has been considered and 
the strong sensitivity of the lineshape to 
the first event in the stochastic trajectory was found.
One of many extensions of this work is to consider 
the random frequency modulation of stochastic oscillator among $N>2$ values, 
and will be published elsewhere.  
This work was supported by NSF.


\end{multicols}
\end{document}